
\input harvmac

\def\np#1#2#3{Nucl. Phys. B{#1} (#2) #3}
\def\pl#1#2#3{Phys. Lett. {#1}B (#2) #3}

\def\physrev#1#2#3{Phys. Rev. {D#1} (#2) #3}

\def\prep#1#2#3{Phys. Rep. {#1} (#2) #3}

\def\cmp#1#2#3{Comm. Math. Phys. {#1} (#2) #3}
\def\ev#1{\langle#1\rangle}

\def\tilde{\widetilde}

\def\CC{{\cal C}}
\def\Tr{{\rm Tr ~}}

\Title{hep-th/9505051, RU-95-27}
{\vbox{\centerline{New RG Fixed Points and Duality in Supersymmetric}
\centerline{$SP(N_c)$ and $SO(N_c)$ Gauge Theories}}}
\bigskip
\centerline{K. Intriligator}
\vglue .5cm
\centerline{Department of Physics and Astronomy}
\centerline{Rutgers University}
\centerline{Piscataway, NJ 08855-0849, USA}

\bigskip

\noindent

We present evidence for new, non-trivial RG fixed points with dual
magnetic descriptions in $N=1$ supersymmetric $SP(N_c)$ and $SO(N_c)$
gauge theories.  The $SP(N_c)$ case involves matter $X$ in the
antisymmetric tensor representation and $N_f$ flavors of quarks $Q$ in the
fundamental representation.  The $SO(N_c)$ case involves matter $X$
in the symmetric tensor representation and $N_f$ flavors of quarks $Q$ in
the vector representation of $SO(N_c)$.  Perturbing these  theories by
superpotentials $W(X)$, we find a variety of interesting RG fixed points
with dual descriptions.  The duality in these theories is similar to that
found by Kutasov and by Kutasov and Schwimmer in $SU(N_c)$ with adjoint
$X$ and $N_f$ quarks in the fundamental.
\Date{5/95}

\lref\ils{K. Intriligator, R.G. Leigh and N. Seiberg, hep-th/9403198,
\physrev{50}{1994}{1092}; K. Intriligator, hep-th/9407106,
\pl{336}{1994}{409}}%
\lref\sv{M.A. Shifman and A. I. Vainshtein, \np{277}{1986}{456};
\np{359}{1991}{571}}
\lref\nonren{N. Seiberg, hep-ph/9309335, \pl{318}{1993}{469}}%
\lref\nati{N. Seiberg, hep-th/9402044, \physrev{49}{1994}{6857}}%
\nref\nonren{N. Seiberg, hep-ph/9309335, \pl{318}{1993}{469}}%
\nref\power{N. Seiberg, The Power of Holomorphy -- Exact Results in 4D
SUSY Field Theories.  To appear in the Proc. of PASCOS 94.
hep-th/9408013, RU-94-64, IASSNS-HEP-94/57}
\nref\ads{I. Affleck, M. Dine and N. Seiberg, \np{241}{1984}{493};
\np{256}{1985}{557}}%
\nref\sv{M.A. Shifman and A. I. Vainshtein, \np{277}{1986}{456};
\np{359}{1991}{571}}
\nref\nsvz{V.A. Novikov, M.A. Shifman, A. I.  Vainshtein and V. I.
Zakharov, \np{223}{1983}{445}; \np{260}{1985}{157};
\np{229}{1983}{381}}%
\nref\cern{D. Amati, K. Konishi, Y. Meurice, G.C. Rossi and G.
Veneziano, \prep{162}{1988}{169} and references therein}%
\lref\om{C. Montonen and D. Olive, \pl {72}{1977}{117}}
\lref\sw{N. Seiberg and E. Witten, hep-th/9407087, \np{426}{1994}{19};
hep-th/9408099, \np{431}{1994}{484}}%
\lref\sem{N. Seiberg, hep-th/9411149 , \np{435}{1995}{129}}%
\lref\isson{K. Intriligator and N. Seiberg, hep-th/9503179, to appear in
Nucl. Phys. B}
\lref\intpou{K. Intriligator and P. Pouliot, hep-th/9505006, RU-95-23}
\lref\kut{D. Kutasov, hep-th/9503121, EFI-95-11}
\lref\kutsch{D. Kutasov and A. Schwimmer, hep-th/9505004, EFI-95-20, WIS/4/95}
\lref\Ahsonyank{O. Aharony, J. Sonnenschein, and S. Yankielowicz,
hep-th/9504113, TAUP-2246-95}
\def\kref{\refs{\kut , \kutsch}}
\lref\rlmsmm{R. Leigh and M. Strassler, hep-th/9503121, RU-95-2}
\newsec{Introduction}

Recent work has shown that $N=1$ supersymmetric gauge theories are a
fruitful arena for studying the dynamics of strongly coupled gauge
theories.  The special feature of these theories is that they contain
holomorphic quantities which can often be obtained exactly \nonren .
See \power\ for a recent review and \refs{\ads - \cern} for earlier
work. One of the most intriguing of the recent
results is the discovery by Seiberg
that $N=1$ theories can have interacting non-Abelian
Coulomb phase fixed points which
can be given dual descriptions in terms of
``magnetic'' gauge theories \sem.  This duality is a generalization of
the Montonen-Olive electric-magnetic duality
\om\ of $N=4$
theories \ref\dualnf{H. Osborn, \pl{83}{1979}{321};
A. Sen, hep-th/9402032,
\pl{329}{1994}{217} }
and $N=2$ theories \sw\ to $N=1$ supersymmetric
theories.   As with the $N=4$ and $N=2$ duality, the $N=1$ duality
exchanges strong and weak coupling effects.  However, in the $N=1$
duality the original electric description and the dual magnetic
description have different
gauge groups and matter content.  In addition, while the $N=4$
and $N=2$ duality are thought to be exact, $N=1$ duality arises only in
the far infrared.  The original
examples of $N=1$ duality
include $SU(N_c)$ with matter in the fundamental
\sem , $SO(N_c)$ with matter in the $N_c$ dimensional vector
representation \refs{\sem , \isson}, and $SP(N_c)$ with matter in the
fundamental \refs{\sem , \intpou}.  Because a general understanding of
duality is not yet known, it is important to have more examples to
gain intuition as to what the phenomenon is.

New examples of non-trivial infrared fixed points with dual descriptions
have recently been discovered by Kutasov and by Kutasov and Schwimmer in
$N=1$ supersymmetric $SU(N_c)$ gauge theory with a single
adjoint field $X$ with a
superpotential $W(X)$ and $N_f$ flavors of quarks \refs{\kut , \kutsch}.
The superpotential $W(X)$ for the adjoint field
plays an important role, controlling the
fixed point to which the theory is driven.  Once a dual description of
a fixed point has been
found, by adding a relevant perturbation we can flow to another
fixed point in the IR with a dual description which is inherited from
that of the initial fixed point. For example, a duality inherited by
perturbing from that of \kut\ was analyzed in \Ahsonyank,
providing a check on the duality of
\kut .  When a fixed point of \kref\ is perturbed to give $X$ a mass, the
duality of \kref\ flows to give the $SU(N_c)$ duality discovered in \sem
.  Perhaps the duality of \kref\ is inherited by
flowing from a dual description of the theory with no
superpotential, though no such dual description is presently known.

In this paper we present evidence for new, non-trivial RG fixed points
with dual magnetic descriptions
in $N=1$ supersymmetric $SP(N_c)$ and $SO(N_c)$ gauge theories.  The
$SP(N_c)$ case, which is discussed in
sect. 2, involves a single matter field $X$ in the
antisymmetric tensor representation with a superpotential $W(X)$
and $N_f$ flavors of quarks in the fundamental.  Perturbing these fixed
points to give $X$ a mass, the duality discussed here flows to the
$SP(N_c)$ case of the duality discussed in \refs{\sem , \intpou}.
The $SO(N_c)$ case, which we discuss in sect. 3, involves
a single matter field $X$ in the symmetric tensor
representation with a superpotential $W(X)$ and $N_f$ flavors of quarks
in the $SO(N_c)$ vector representations.  Perturbing these fixed points
to give $X$ a
mass, the duality discussed here flows to the $SO(N_c)$ case of the
duality discussed in \refs{\sem ,\isson}.
The duality in these theories is very similar to that of \kref\ though,
unlike the $SU(N_c)$ case, the matter field $X$ is here not in the
adjoint representation.
The analog of \kref\ for $SO(N_c)$ and
$SP(N_c)$ with adjoint matter will appear in \ref\msrlad{M. Strassler
and R. Leigh, Rutgers preprint to appear}.
To stress the similarity of the duality discussed here with that of
the $SU(N_c)$ case, our analysis will closely follow that of \kutsch .

\newsec{$SP(N_c)$ with an antisymmetric and $2N_f$ fundamentals}

We consider $SP(N_c)$\foot{$SP(N_c)$ is the subgroup of
$SU(2N_c)$ which leaves invariant an antisymmetric tensor $J^{ab
}$, which we can take to be $J={\bf 1}_{N_c}\otimes i\sigma _2$.}
with matter $X$ in the $(N_c(2N_c-1)-1)$ dimensional ``traceless''
antisymmetric tensor representation of $SP(N_c)$, $X_{ab}=-X_{ba}$
and $X^a_a\equiv J^{ab}X_{ba}=0$,
and $2N_f$ fields $Q_f$, $f=1\dots 2N_f$, in the $2N_c$ dimensional
fundamental representation of $SP(N_c)$.
This theory is asymptotically free for $N_f\leq 2N_c+4$ and, for
sufficiently large $N_f$, has an interacting, non-Abelian
Coulomb phase fixed point in the infrared.
Rather than analyze this theory, it is
easier to instead consider the theory perturbed by the
superpotential
\eqn\wspi{W=g_k\Tr X^{k+1},}
where color indices are contracted with $J^{ab}$.
For $k>2$ the superpotential \wspi\ looks irrelevant near the UV fixed
point.  Nevertheless, there is a range of $N_f$ depending on $k$ for
which the superpotential \wspi\ is actually relevant in the IR, driving
the theory to a new fixed point.

The theory with superpotential \wspi\ has an anomaly free global
$SU(2N_f)\times U(1)_R$
symmetry with the matter fields transforming as
\eqn\spegt{\eqalign{Q&\qquad (2N_f, 1-{2(N_c+k)\over (k+1)N_f})\cr
X&\qquad (1, {2\over k+1}).}}
The gauge invariant, non-redundant operators of the theory are $\Tr
X^{j-1}$ and generalized ``mesons''
$(M_j)_{fg}=Q_fX^{j-1}Q_g$, for $j=1\dots k$.  Each $M_j$ is in the
$N_f(2N_f-1)$ dimensional antisymmetric representation of $SU(2N_f)$.
There are no ``baryons'' in $SP(N_c)$;
because
$\epsilon ^{a_1\dots a _{2N_c}}$ breaks
up into sums of products of the $J^{ab}$, baryons break up into
mesons.

\subsec{Stability}

Consider deforming \wspi\ to include lower order terms:
\eqn\wspid{W=\Tr\sum _{l=1}^kg_lX^{l+1}+\lambda \Tr X,}
where $\lambda$ is a Lagrange multiplier to enforce $\Tr X\equiv
J^{ab}X_{ba}=0$.
The theory has multiple vacua with $\ev{Q}=0$ and $\ev{X}\neq 0$,
satisfying $W'(X)=0$.
Because $X$ is antisymmetric, the eigenvalues of $JX$ come in pairs $x_l$,
satisfying $W'(x_l)=0$.  Because $W'(x)$ is of degree $k$, there are
$k$ solutions $x_i$.  Let $i_l$ be the number of pairs of eigenvalues of $JX$
equal to $x_l$, with $\sum _{l=1}^ki_l=N_c$.  In such a vacuum the
gauge group is broken by $\ev{X}$ as:
\eqn\sphiggs{SP(N_c)\rightarrow SP(i_i)\times SP(i_2)\times \cdots
\times SP(i_k).}
In each vacuum $X$ is massive and can be integrated out.  Each $SP(i_l)$
factor has $N_f$ flavors.  As discussed in \intpou, $SP(N_c)$ with
$N_f\leq N_c$ does not have a stable vacuum.  A special case for the
theories with stable vacua is $N_f=N_c+1$, which does does not have a
vacuum at the origin.
Therefore, in order for
our theory to have a stable vacuum, we must have
\eqn\spstabii{\eqalign{i_l&<N_f\qquad\hbox{for all}\ l=1\dots k\cr
\hbox{and}\quad i_l+1&<N_f\qquad\hbox{for a vacuum at the origin}.}}
Our original theory \wspi\ thus has a stable vacuum provided
\eqn\spstab{N_f>{N_c\over k}.}

\subsec{Duality}

The dual magnetic description of the fixed point of the theory with
superpotential \wspi\ is in terms of an
$SP(\tilde N_c)$ theory, where $\tilde N_c\equiv
k(N_f-2)-N_c$, with matter $Y$ in the
antisymmetric traceless representation, $N_f$ flavors of quarks $q^f$ in the
fundamental and gauge singlets $(M_j)_{f,g}=-(M_j)_{g,f}$, $j=1\dots k$
and $f,g=1, \dots 2N_f$, with superpotential
\eqn\wdspi{W=\Tr Y^{k+1}+\sum _{j=1}^kM_jqY^{k-j}q.}
(There are constants $C_j$ implicit in the
$j$-th term in \wdspi .  Dualizing again will yield the original
electric theory provided the $\tilde C_j$ in
the dual of the dual satisfy $\tilde C_j=-C_{k+1-j}$.)
As in \refs{\sem , \kut, \kutsch}, the operators $Q_fX^{j-1}Q_g$ of the
electric theory are represented by the gauge singlet fields $(M_j)_{fg}$
in the dual theory.  The operators $\Tr X^j$, $j=2\dots k$, are mapped
to $\Tr Y^j$ in the dual theory.

Taking $M_j$ to transform as $QX^{j-1}Q$, the dual
theory has a global $SU(2N_f)\times U(1)_R$ symmetry with the fields
transforming as
\eqn\spmgt{\eqalign{q&\qquad (\overline{2N_f},
1-{2(\tilde N_c+k)\over (k+1)N_f})\cr
Y&\qquad (1,{2\over k+1})\cr
M_j&\qquad (N_f(2N_f-1),2{k+j\over k+1}-{4(N_c+k)\over (k+1)N_f}).}}
Note that this
symmetry is anomaly free in the dual $SP(\tilde N_c)$ theory.
At the origin of the space of flat directions the $SU(2N_f)\times
U(1)_R$ symmetry is unbroken and the 't
Hooft anomalies computed with the original spectrum of the electric
theory must match those computed with the dual magnetic spectrum.  It is
a highly non-trivial check of the duality
that they do indeed match; both spectra give
\eqn\spthoofti{\eqalign{U(1)_R\qquad &-{2N_c\over
k+1}(2N_c+3k)+{k-1\over k+1}\cr
U(1)_R^3\qquad &-{32N_c(N_c+k)^3\over N_f^2(k+1)^3}+N_c(2N_c+1)-
{(k-1)^3(2N_c^2-N_c-1)\over (k+1)^3}\cr
SU(2N_f)^3\qquad &2N_cd_3(2N_f)\cr
SU(2N_f)^2U(1)_R\qquad &-{4N_c(N_c+k)\over (k+1)N_f}d_2(2N_f),}}
where $d_2(2N_f)$ and $d_3(2N_f)$ are the quadratic and cubic $SU(2N_f)$
Casimirs in the fundamental representation.

\subsec{Deformations}

We can deform our fixed points either by perturbing the superpotential or by
giving expectation values to some fields along the $D$-flat directions.
Any perturbation of the electric theory must have a corresponding dual
perturbation in the magnetic theory which must generate a RG flow which is
dual to the electric RG flow.   In particular, the new low energy fixed
point will have a duality
which is inherited from that of the original fixed point.  Checking that
the electric and magnetic flows and, in particular, the new fixed points
really are dual provides highly non-trivial checks on the duality.  We
will briefly discuss a variety of perturbations, checking the duality.
\medskip
\centerline{\it Superpotential deformations}
\medskip
We first consider deforming the theory by giving a mass to one flavor of the
electric quarks.  In the electric theory the tree level superpotential is
\eqn\wspelm{W_{\rm elec}=g_k\Tr X^{k+1}+mQ_{2N_f-1}Q_{2N_f}.}
The low energy theory is an $SP(N_c)$ theory with matter $X$ and
superpotential \wspi\ with one fewer flavor, $\hat N_f=N_f-1$.  The low
energy theory, having fewer flavors, is stronger in the infrared.
In the dual theory this perturbation corresponds to
\eqn\wspmm{W_{\rm mag}=g_k\Tr Y^{k+1}+
\sum_{j=1}^k M_j q Y^{k-j}q+m(M_1)_{2N_f-1,2N_f}.}
The $M_j$ equations of motion imply that the vacua of this theory satisfy
\eqn\speom{q^{2N_f-1}Y^{l-1}q^{2N_f}=-\delta_{l,k}m;\;\;l=1,\cdots, k}
which, along with some additional conditions, give expectation
values proportional to:
\eqn\spmv{\eqalign{ q^{2N_f-1}_c=&\delta_{c,1};\cr q^{2N_f}_c=
&\delta_{c,2k};\cr
Y_{c,d}=&\cases{\delta_{c+1,d}&$c=2r;\ r=1,\cdots, k-1$\cr
-\delta_{d+1, c}&$d=2r;\ r=1,\cdots, k-1$\cr
0& otherwise.\cr}\cr}}
These expectation values break the magnetic $SP(k(N_f-2)-N_c)$ gauge
group to $SP(k(N_f-3)-N_c)$ with $N_f-1$ remaining light flavors.
The low energy magnetic theory is at weaker coupling and is the dual of
the low energy electric theory.

We can also consider perturbing by other $M_l$.  Consider, for example,
perturbing the electric theory by adding to \wspi\ a term
$h_rQ_{2N_f-1}X^{r-1}Q_{2N_f}$.  In
the magnetic description the superpotential is \wdspi\ with an additional
term $h_r(M_r)_{2N_f-1,2N_f}$.  The $M$ equations of motion give
$q^{2N_f-1}Y^{k-r}q^{2N_f}=-h_r$, breaking the dual gauge group to
$SP(k(N_f-3)+r-1-N_f)$.  The low energy theory has a duality inherited
{}from the duality discussed here.

Another type of superpotential perturbation is by $\Tr X^r$.  As in
the $SU(N_c)$ case \refs{\Ahsonyank ,\kutsch},
the vacuum stability plays an important role in verifying that
the duality works.
Consider, for example, deforming the $k=2$ case of \wspi\ by a mass term
for the field $X$
\eqn\wkiispm{W_{\rm elec}=\Tr (X^3+\half mX^2+\lambda X).}
The quadratic equation $W'=0$ for the eigenvalues has solutions $x_\pm$.
There are vacua with $r$ eigenvalue pairs equal to $x_+$ and $N_c-r$
equal to $x_-$ for $r=0, \dots N_c$.  In such a vacuum the gauge group
is broken as
\eqn\kiispme{SP(N_c)\rightarrow SP(r)\times SP(N_c-r);}
$X$ is massive and each factor has $N_f$ flavors of $Q$.
Taking $N_f>N_c$, each factor in \kiispme\ satisfies \spstabii\ for
all $r=0 \dots N_c$ and thus all $N_c+1$ vacua are stable.

In the dual theory a similar analysis gives vacua labeled by
$\tilde r=0\dots 2(N_f-2)-N_c$ with the magnetic gauge group broken as
\eqn\kiispmm{SP(2(N_f-2)-N_c)\rightarrow SP(\tilde r)\times
SP(2(N_f-2)-N_c-\tilde r).}
$Y$ is massive and each factor has $N_f$ flavors of $q$ along with gauge
singlets $M$, coming from linear combinations of
$M_1$ and $M_2$, with a $W=Mqq$ superpotential.
The $M$ equation of motion and the D-terms give $q=0$; therefore, there
is a vacuum provided each dual theory has a vacuum at $q=0$.  This
requires  $\tilde r+1<N_f$ and
$2(N_f-2)-N_c-\tilde r+1<N_f$, which gives $N_c+1$ values of
$\tilde r$ with stable vacua, as required by the duality.
The map between the factors in \kiispme\ and \kiispmm\ is the duality of
\refs{\sem ,\intpou}, $\tilde r=N_f-2-r$.

\medskip
\centerline{\it Flat direction deformations}

We can consider deforming the theory along the flat directions with
various $\ev{M_j}\neq 0$.  Consider, for example, the flat direction
rank$(\ev{M_1})=2$ with $\ev{M_{j>1}}=0$, corresponding to the
expectation value of a single flavor, $\ev{Q_{2N_f-1}Q_{2N_f}}\neq 0$,
with $\ev{X}=0$.
Along this flat direction the electric gauge group is broken to
$SP(N_c-1)$ with $N_f-1$ light $Q_{\hat f}$, $\hat f=1\dots
2(N_f-1)$.  In addition, there are
two more $SP(N_c-1)$ fundamentals $F_{1,2}$, a
singlet $S$, and an $SP(N_c-1)$ antisymmetric $\hat X$, all coming from
$X$; these fields have interactions inherited from $W=\Tr X^{k+1}$.  The
electric theory is at weaker coupling.

In the dual magnetic description, the above flat direction corresponds to a
large term $\ev{M_{2N_f-1,2N_f}}q^{2N_f-1}Y^{k-1}q^{2N_f}$ in the
superpotential.  The low energy magnetic theory is at stronger coupling
and is the dual description of the low energy electric theory with the
fields mentioned above and the superpotential inherited from
$W=X^{k+1}$.  Giving $\ev{M_1}$ larger rank, the dual magnetic theory
must cease to have a vacuum when rank$(\ev{M_1})\geq N_c$ in order to
reproduce what is a classical consequence of $M_1=QQ$ in the electric
theory.

For $N_c=kn$ the theory \wspi\ also has flat directions with
$\ev{X}\neq 0$ and $\ev{Q}=0$.  Along these flat directions the electric
gauge group is broken as
\eqn\spxfd{SP(N_c)\rightarrow SP(n)^k,}
with $N_f$ flavors in each $SP(n)$.  In the dual theory this flat
direction corresponds to the flat direction $\ev{Y}\neq 0$, $\ev{q}=0$.
Along this flat direction the magnetic gauge group is broken as
\eqn\spyfd{SP(\tilde N_c)\rightarrow SP(N_f-2-n)^k;}
in each $SP(N_f-2-n)$ theory there are
$N_f$ flavors of quarks $q$ and gauge singlets $M$, which are linear
combinations of the $M_j$, coupled with
$W=Mqq$.  Along this flat direction, the duality flows to $k$ copies
of the duality discussed in \refs{\sem , \intpou}.

\newsec{$SO(N_c)$ with a symmetric and $N_f$ vectors}

We now consider $SO(N_c)$ with matter $X$ in the $(\half N_c(N_c+1)-1)$
dimensional symmetric traceless tensor representation of $SO(N_c)$,
$X_{cd}=X_{dc}$ with $X_{cd}\delta ^{cd}=0$,
and $N_f$ fields $Q_f$, $f=1\dots N_f$, in the
$N_c$ dimensional vector representation
of $SO(N_c)$.  This theory is asymptotically free for
$N_f\leq 2(N_c-4)$, with an interacting non-abelian Coulomb phase in the
infrared.  We consider the theory deformed by the superpotential
\eqn\wsoi{W=g_k\Tr X^{k+1}.}
Again, for a range of $N_f$ depending on $k$, \wsoi\ becomes relevant in
the infrared, driving the theory to a new fixed point.

The gauge invariant, non-redundant operators include the
$\Tr X^{j-1}$ and the $M_j=QX^{j-1}Q$, $j=1\dots k$.  The $M_j$ are all in
the $\half N_f(N_f+1)$ dimensional symmetric representation of
$SU(N_f)$. Additional gauge
invariant, non-redundant operators can be made by contracting
gauge indices with an $\epsilon$ tensor:
\eqn\soexops{
B_p^{(n_1, \dots ,n_k)}=(W_\alpha )^pQ_{(1)}^{n_1}...Q_{(k)}^{n_k};
\qquad \sum_{l=1}^kn_l =N_c-2p,}
where $Q_{(l)}\equiv X^{l-1}Q$.  For odd $p$, $B_p$
transform with a
Lorentz spinor index $\alpha$ while, for even $p$, $B_p$ are scalar chiral
superfields.  The total number of operators of the form \soexops\ is
\eqn\ttnm{\sum_{\{n_l\}}{N_f\choose n_1}\cdots{N_f\choose n_k}=
{kN_f\choose N_c-2p}.}

The theory with \wsoi\ has an anomaly free global
$SU(N_f)\times U(1)_R$ symmetry with matter fields transforming as
\eqn\soegt{\eqalign{Q&\qquad (N_f,1-{2(N_c-2k)\over (k+1)N_f} )\cr
X&\qquad (1, {2\over k+1}).}}
In addition, the theory is invariant under the discrete $Z_{2N_f}$
symmetry generated by $Q\rightarrow e^{{2\pi i\over 2N_f}}Q$ and also
charge conjugation $\CC$.

\subsec{Stability}

As discussed in \isson , $SO(N_c)$ with $N_f$ vectors has a stable
vacuum provided $N_f\geq N_c-4$.  Repeating the discussion following
\wspid , the expectation value $\ev{X}$ breaks the gauge group as
\eqn\soxhigs{SO(N_c)\rightarrow SO(i_1)\times SO(i_2)\times \cdots
SO(i_k).}
In each vacuum $X$ is massive and can be integrated out.
Every $SO(i_l)$ has $N_f$ vectors and, therefore,
there is a stable vacuum provided
\eqn\sostabvi{i_l-4\leq N_f\qquad\hbox{for all}\ l=1\dots k.}
Therefore, the theory \wsoi\ has a stable vacuum provided
\eqn\sostabii{N_f\geq {(N_c-4)\over k}.}

\subsec{Duality}

The dual theory is an $SO(\tilde N_c)$ theory, where $\tilde N_c\equiv
k(N_f+4)-N_c$, with matter $Y$ in the symmetric traceless tensor
representation of $SO(\tilde
N_c)$, $N_f$ fields $q^f$ in the
vector representation of $SO(\tilde N_c)$, singlets
$(M_j)_{fg}=(M_j)_{gf}$ with $j=1\dots
k$ and $f,g, =1 \dots N_f$, and a superpotential
\eqn\wsodi{W=\Tr Y^{k+1}+\sum _{j=1}^kM_jqY^{k-j}q.}

Taking $M_j$ to transform as $QX^{j-1}Q$, the dual
theory has a global $SU(N_f)\times U(1)_R$ symmetry with fields
transforming as
\eqn\spmgt{\eqalign{q&\qquad (\overline{2N_f},
1-{2(\tilde N_c-2k)\over (k+1)N_f})\cr
Y&\qquad (1,{2\over k+1})\cr
M_j&\qquad (\half N_f(N_f+1),{2(j+k)\over k+1}-{4(N_c-2k)\over
(k+1)N_f}).}}
Note that this
symmetry is anomaly free in the dual $SO(\tilde N_c)$ theory.
The discrete symmetries,
generated by $Q\rightarrow e^{{2\pi i\over 2N_f}}Q$ and charge
conjugation $\CC$ in the electric theory, are generated by
$q\rightarrow e^{-{2\pi i\over 2N_f}}\CC ^kq$ and $\CC$, respectively,
in the dual theory.

It is a non-trivial check on the duality that the 't
Hooft anomalies computed with the original electric spectrum and the dual
magnetic spectrum anomalies match; both give
\eqn\spthoofti{\eqalign{U(1)_R\qquad &{N_c(-N_c+3k)\over k+1}+{k-1\over
k+1}\cr
U(1)_R^3\qquad &-{8N_c(N_c-2k)^3\over N_f^2(k+1)^3}+\half N_c(N_c-1)-
{(k-1)^3(N_c^2+N_c-2)\over 2(k+1)^3}\cr
SU(N_f)^3\qquad &N_cd_3(N_f)\cr
SU(N_f)^2U(1)_R\qquad &-{2N_c(N_c-2k)\over (k+1)N_f}d_2(N_f).}}

It is also non-trivial that there exists a mapping between the operators
\soexops\ and the corresponding operators in the dual theory which is
consistent with all of the global continuous and descrete symmetries.
Such a mapping is
\eqn\exopmp{B_p^{(n_1, n_2, \dots , n_k)}\leftrightarrow (\Tr
Y^{(k-1)(k-p)})
\tilde B
_{\tilde p}^{(\tilde n_1, \tilde n_2, \dots, \tilde n_k)};\quad \tilde
n_l=N_f-n_{k+1-l},} with $\tilde p=2k-p$.

For $k=1$ the field $X$ is massive and can be integrated out.  In this
case, the duality discussed here reduces to the duality discussed in
\refs{\sem , \isson}.

\subsec{deformations}
As in sect. 2.3, we will briefly discuss a few deformations of the fixed
points, checking the duality.
\medskip
\centerline{\it Superpotential deformations}
\medskip

We first consider deforming the electric theory by adding a mass term
for the $N_f$-th quark.  In the magnetic theory the superpotential is
similar to \wspmm .  The vacuum has
\eqn\soeom{q^{N_f}Y^{l-1}q^{N_f}=-m\delta _{l,k};\ l=1,
\dots , k,}
which give expectation values $\ev{Y}$ and $\ev{q}$,
breaking $SO(k(N_f+4)-N_c)$ with $N_f$
quark flavors to $SO(k(N_f+3)-N_c)$ with $N_f-1$ quark flavors.  For
example, for $k=2$ the expectation values are proportional to
$Y_{11}=-Y_{22}=iY_{12}=iY_{21}=1$, with all other components zero, and
$q^{N_f}_{c}=\delta _{c,1}+i\delta _{c,2}$, which reduces the number of
colors by two. The low energy magnetic theory is the dual
of the low energy electric theory.

We consider deforming the $k=2$ case by a mass term for the field
$X$,
\eqn\wkiisom{W_{\rm elec}=\Tr\ (X^3+\half mX^2+\lambda X).}
Again, the equation $W'(X)=0$ for the eigenvalues has solution $x_\pm$
and there are vacua with $r$ eigenvalues of $X$ equal to $x_+$ and
$N_c-r$ equal to $x_-$ for $r=0, \dots , N_c$.  In such a vacuum the
gauge group is broken as
\eqn\kiisome{SO(N_c)\rightarrow SO(r)\times SO(N_c-r).}
In each factor $X$ is massive and there are $N_f$ flavors of $Q$.
Taking $N_f\geq N_c-4$, each factor satisfies \sostabvi ;
there are $N_c+1$ stable vacua corresponding to $r=0, \dots , N_c$.

In the dual theory a similar analysis gives vacua labeled by
$\tilde r=0, \dots , 2(N_f+4)-N_c$ with the magnetic gauge group broken
as
\eqn\kiisomm{SO(2(N_f+4)-N_c)\rightarrow SO(\tilde r)\times
SO(2(N_f+4)-N_c-\tilde r).}
In each vacuum $Y$ is massive and there are $N_f$ flavors of $q$ coupled
to gauge singlets $M$ with a superpotential $W=Mqq$.
The $M$ equations of motion and the $D$
terms fix $q=0$.  The dual theory has a vacuum there provided $\tilde
r-4\leq N_f$ and $2N_f+4-N_c-\tilde r\leq N_f$. There are $N_c+1$ such
values of $\tilde r$, corresponding to the $N_c+1$ vacua of the electric
theory.  The duality map between each factor in \kiisome\ and \kiisomm\
is as in \refs{\sem, \isson}: $\tilde
r=N_f+4-r$.  The vacua with $r=0$ and $r=N_f$
are doubly degenerate, corresponding to
gaugino condensation in the magnetic theory.  The others have massless
fields, corresponding to the original massless electric quarks.  For
example, for $r=2$ one of the components of \kiisome\ is an $SO(2)$
gauge theory with $N_f$ massless charged quarks $Q$.  The corresponding
magnetic theory is $SO(N_f+2)$ with $N_f$ flavors and the superpotential
which fixes $q=0$.  As in \isson , the magnetic theory at $q=0$ has
$N_f$ collective excitations $\tilde q_f$
which are magnetically charged relative to the magnetic gauge group.
These are the electrically charged components of the
quarks of the electric theory.
\medskip
\centerline{\it Flat direction deformations}
\medskip

We can consider deforming the electric theory by taking various
$\ev{M_j}\neq 0$.  These expectation values must satisfy various
relations, which are classical in the electric theory, and take the
low energy theory to weaker coupling.  In the magnetic theory,
the $\ev{M_j}$ act
as ``generalized mass'' terms in the superpotential
for magnetic quarks, taking the low energy
theory to stronger coupling.  The low energy electric and magnetic
theories have a duality which is inherited from that discussed here.

For $N_c=kn$ the electric theory \wsoi\ also has vacua with $\ev{X}\neq 0$
and $\ev{Q}=0$.  In these vacua $SO(N_c)$ is broken to $SO(n)^k$ with
$N_f$ flavors in each factor.  The corresponding flat direction in the
dual theory breaks the magnetic gauge group to $SO(N_f+4-n)^k$
with $N_f$ flavors $q$ and a meson $M$ with superpotential $W=Mqq$
in each factor.  The duality inherited in each factor is that discussed
in \refs{\sem , \isson}.

\newsec{Conclusions}

As in \kref, the fixed points and duality discussed here are limited to
the theories with the superpotential $W(X)\neq 0$.  It is an interesting
open problem to understand the fixed points of the $W=0$ theory
and perhaps the duality from which ours is inherited.  As in
\kutsch , our results suggest the following picture for the theories
with $W=0$:  As $N_f$ decreases from $D$, where $D=2N_c+4$ for $SP(N_c)$
and $D=2(N_c-4)$ for $SO(N_c)$, $X^{k+1}$ becomes less irrelevant in the
infrared until, for $N_f$ less than some $N_0(k)$, $X^{k+1}$ becomes
relevant in the infrared.  For $N_f<N_0(k)$, the theories are driven to the
new fixed points.  Also, as in \kutsch , gauge invariant
operators such as $\Tr X^2$ must decouple when $N_f$ decreases to the point
when their dimensions drop down to one.  In this way, the fixed points
which we consider can evade what looks like a problem
with the unitarity bounds \ref\mack{G. Mack, \cmp{55}{1977}{1}} for
$k>5$ (where the dimension of $\Tr X^2$, were it to be related to its
$R$ charge, drops below one).  These issues deserve further
investigation.

It has been suggested in \refs{\sem ,\rlmsmm}
that the duality of \sem\ could be related
to the duality of ``finite'' $N=2$ theories.  Similarly, it was
suggested in \kutsch\ that the duality discussed there is related to
$N=2$ duality.
The duality discussed here is harder to relate to $N=2$
duality.  For example, there is no asymptotically free $N=2$ $SO(N_c)$
theory with a symmetric tensor hypermultiplet.
It is, however, amusing to note that the field $X$ of $SP$ is
similar to an $SO$ adjoint and vice-versa and that these groups
are magnetic groups \ref\ivsosp{P. Goddard, J. Nuyts, and D. Olive,
\np{125}{1977}{1}} of each other (for example, they are exchanged
in $N=4$ duality).
I hope that these examples will be useful in illuminating the nature
of duality.

\centerline{{\bf Acknowledgments}}

I would like to thank R. Leigh, P. Pouliot, N. Seiberg, and M. Strassler
for discussions.  This work was supported in part by DOE grant
\#DE-FG05-90ER40559

\listrefs
\end